\documentclass[preprint]{revtex4}
\usepackage{amsmath}
\include{graphics}
\begin{document}
\bibliographystyle{apsrev}
\input{epsf}
\title{Equivalence of Two Approaches for Quantum-Classical Hybrid Systems}

\author{Fei Zhan}
\affiliation{Institute of Physics, Chinese Academy of Sciences,
Beijing 100190, China}

\author{Yuan Lin}
\affiliation{Institute of Physics, Chinese Academy of Sciences,
Beijing 100190, China}

\author{Biao Wu}
\affiliation{Institute of Physics, Chinese Academy of Sciences,
Beijing 100190, China}

\begin{abstract}
We discuss two approaches that are used frequently to describe
quantum-classical hybrid system.  One is the well-known
mean-field theory and the other adopts a set of hybrid brackets
which is a mixture of quantum commutators and classical Poisson
brackets.  We prove that these two approaches are equivalent.
\end{abstract}
\date{\today}
\maketitle
Other than systems that are either fully quantum or fully
classical, there are many hybrid systems, where a quantum subsystem is
coupled to a classical subsystem.  These quantum-classical hybrid systems
are important and interesting to researchers from very different backgrounds.
Since gravity has not been properly quantized, it has been pondered whether it
is ever necessary to quantize gravity\cite{Rosenfeld1963NPhy,Kibble1981}.
If not, then one has to deal with
hybrid systems where classical gravity is coupled to other quantized
field\cite{ref:andsn}. Hybrid systems are also encountered in quantum
measurement, where the detector, which is coupled to a quantum system,
is always classical\cite{Braginsky1992Book,ref:gisin}. On a more practical side,
hybrid systems are also studied by researchers who are interested in
the properties of solids and molecules. Even these systems are fundamentally
quantum, it is adequate to treat the electrons as quantum
while treating the heavy ions as classical\cite{ref:delos1,ref:delos2,ref:tully489,ref:pk,ref:prezhdomean}.
This is, of course, the well-known Born-Oppenheimer approximation\cite{Born1927}.

Because of the diversity of people who are interested in hybrid
systems, various different approaches have been proposed to these
half quantum and half classical systems. These approaches include
the mean-field theory \cite{ref:eh,ref:prezhdomean,ref:zq},
quasiclassical bracket approach
\cite{ref:alek,ref:kapral,ref:nielsen}, Bohmian
method\cite{ref:prezhdoback,ref:ginden}, decoherent histories
\cite{ref:hallid,ref:bittner}, and many others
\cite{ref:tullytrajectory,ref:tullymolecular,ref:jonesd}. Among
these approaches, the most popular ones are the mean-field theory
and the quasiclassical brackets.  In the mean-field theory, the
quantum subsystem evolves according to the Schr\"odinger equation
while the classical subsystem experiences an energy field which is
the expectation value of the quantum state. In the quasiclassical
bracket approach,  brackets that are mixtures of quantum commutators
and classical Poisson brackets are introduced and used to derive the
equations of motion of the hybrid system.

In this paper we prove that the mean-field theory and the quasiclassical
bracket approach are equivalent.  Before we proceed to present our
proof, we briefly introduce these two approaches.

The Hamiltonian of a hybrid system where there is interaction
between subsystems has three parts: the quantum mechanical
part $\hat{H}_q$, the classical part $H_c$, and the interaction
part $\hat{H}_i$. Formally, the Hamiltonian can be written as
\begin{equation}
\hat{H}=\hat{H}_q(\hat{\textbf{p}},\hat{\textbf{q}})+H_c(\textbf{P},\textbf{Q})+
\hat{H}_i(\hat{\textbf{q}},\textbf{Q}),
\end{equation}
where the dependence of $\hat{H}_i$ only on the coordinates $\textbf{q}$
and $\textbf{Q}$ reflects most cases in physical problems.

In the mean-field theory, one uses the following Hamiltonian
\begin{equation}
\label{mf}
H_s=\langle\psi|\hat{H}_q(\hat{\textbf{p}}_s,\hat{\textbf{q}}_s)+
\hat{H}_i(\hat{\textbf{q}}_s,\textbf{Q}_s)|\psi\rangle+
H_c(\textbf{P}_s,\textbf{Q}_s)\,,
\end{equation}
where $|\psi\rangle$ is a wavevector describing the quantum subsystem.
The subscript $s$ is introduced to distinguish the dynamical variables in
the mean-field approach to the same variables in the quasiclassical
bracket approach that is to be discussed later.  Because the quantum
system possesses mathematically the classical Hamiltonian structure
\cite{ref:heslot,ref:weinberg,ref:lj,ref:zq}, we
introduce the following set of Poisson brackets,
\begin{eqnarray}
&&\{\psi^*_j,\psi_k\}=i\delta_{jk}/\hbar\,,~~~\{Q_j,P_k\}=\delta_{jk}\,,\\
&&\{\psi_j,\psi_k\}=\{Q_j,Q_k\}=\{P_j,P_k\}=0\,,
\end{eqnarray}
where $\psi_j$ is the $j$th component of the wavevector $|\psi\rangle$
when it is expanded in an orthonormal basis,
\begin{equation}
|\psi\rangle=\sum_j\psi_j|j\rangle\,.
\end{equation}
With these Poisson brackets, one can derive a set of equations
of motion from the mean-field Hamiltonian in Eq.(\ref{mf})
\begin{align}
\label{sch}
|\dot{\psi}\rangle=&\frac{1}{i\hbar}\big[\hat{H}_{qs}+\hat{H}_{is}\big]|\psi\rangle\,,\\
\dot{\textbf{Q}}_s=&\frac{\partial H_{s}}{\partial \textbf{P}_s}
=\frac{\partial H_{cs}}{\partial \textbf{P}_s}\,,\label{qs}\\
\dot{\textbf{P}}_s=&-\frac{\partial H_{s}}{\partial
\textbf{Q}_s}=-\frac{\partial }{\partial \textbf{Q}_s}
\Big[\langle\psi|\hat{H}_{is}|\psi\rangle+H_{cs}\Big]\,.\label{ps}
\end{align}
where $\hat{H}_{qs}$ is a shorthand notation for
$\hat{H}_q(\hat{\textbf{p}}_s,\hat{\textbf{q}}_s)$
and similarly for $\hat{H}_{is}$ and $H_{cs}$. The mean field force in Eq.(\ref{ps})
\begin{equation}
\label{fs}
{\bf F}_s=-\frac{\partial }{\partial
\textbf{Q}_s}\Big[\langle\psi|\hat{H}_{is}|\psi\rangle\Big]\,,
\end{equation}
is just the Hellman-Feynman force has been used widely in molecular
dynamic simulations\cite{ref:ep}.

Note that the equations of motion in Eqs.(\ref{sch})-(\ref{ps}) are
usually written down directly
\cite{ref:delos1,ref:delos2,ref:hammes,ref:prezhdomean,ref:hallid,ref:dh}.
To derive them in a coherent theoretical framework as we have
presented was first done in Ref.\cite{ref:zq}.

The quasiclassical bracket approach explores the similarity between
classical Poisson brackets and quantum commutators. In this approach,
one tries to find the equations of motion for  hybrid systems by introducing
a new set of brackets which are mixtures of classical Poisson brackets
and quantum commutators. We call these new brackets quasiclassical
brackets, a name used by Anderson\cite{ref:andsn}.  There are several
different kinds of quasiclassical brackets
\cite{ref:andsn,ref:alek,ref:boucher,ref:pk,ref:kisil,ref:prezhdoquantum}.
One of these quasiclassical  brackets is\cite{ref:andsn}
\begin{equation}
\label{qc}
[A, B]_{qc}=[A, B]+i\hbar\{A, B\}.
\end{equation}
The difference between these different quasiclassical brackets is
subtle\cite{ref:diosi,ref:salcedo}. However, this subtle difference disappears
for most of the interesting
systems in physics, whose Hamiltonian contains no  terms that are
multiples of non-commutative operators. For example, there are no terms like
$\hat{\bf{q}}\cdot\hat{\bf{p}}$ in the Hamiltonians for almost all
systems in nature.  In this paper, we consider only this class of systems
and use the bracket in Eq.(\ref{qc}) to avoid controversy or confusion.

In this quasiclassical bracket approach, the Hamiltonian is different
from the one in the mean-field theory; we write it as
\begin{equation}
\hat{H}_h=\hat{H}_q(\hat{\textbf{p}}_h,\hat{\textbf{q}}_h)+
\hat{H}_i(\hat{\textbf{q}}_h,\textbf{Q}_h)+H_c(\textbf{P}_h,\textbf{Q}_h),
\end{equation}
where the subscript $h$ is the counterpart of the subscript $s$ in the mean-field
theory. With the quasiclassical brackets in Eq.(\ref{qc}), we can obtain
a set of Heisenberg-like equations of motion
\begin{align}
\dot{\hat{\textbf{q}}}_h(t)=&\frac{1}{i\hbar}\left[\hat{\textbf{q}}_h(t),\hat{H}_{qh}(t)+\hat{H}_{ih}(t)\right]_{qc}\,,\label{qqh}\\\
\dot{\hat{\textbf{p}}}_h(t)=&\frac{1}{i\hbar}\left[\hat{\textbf{p}}_h(t),\hat{H}_{qh}(t)+\hat{H}_{ih}(t)\right]_{qc}\,,\label{pph}\\
\dot{\textbf{Q}}_h(t)=&\frac{\partial H_{ch}}{\partial
\textbf{P}_h},\label{qh}\\
\dot{\textbf{P}}_h(t)=&-\langle t_0|\frac{\partial }{\partial
\textbf{Q}_h}\hat{H}_{ih}(t)|t_0\rangle-\frac{\partial }{\partial
\textbf{Q}_h}H_{ch},\label{ph}
\end{align}
where we have used shorthand notations $\hat{H}_{qh}=
\hat{H}_{q}(\hat{\textbf{q}}_h,\textbf{Q}_h)$, $\hat{H}_{ih}=
\hat{H}_{i}(\hat{\textbf{q}}_h,\textbf{Q}_h)$, and $H_{ch}=
H_{c}(\textbf{P}_h,\textbf{Q}_h)$. The wavevector $|t_0\rangle$ is
the initial wavevector of the quantum subsystem. One of the many
problems for the quasiclassical brackets is that one may have to
deal with equations whose left hand side is a $c$-number while whose
right hand side is an operator\cite{ref:jones}. To overcome this,
we have taken the expectation value of the right hand side over the
initial wavevector in Eq.(\ref{ph}) as in
Ref.\cite{ref:prezhdoquantum}.

We now set to prove that the mean-field theory and the
quasiclassical approach to hybrid systems are equivalent, that is,
the dynamics described by the set of equations
Eqs.(\ref{sch})-(\ref{ps}) is the same as the one by
Eqs.\eqref{qqh}-\eqref{ph}.

% After calculation of the set of equations, expectation values of any
% quantum observable can be got from the calculation of $\langle\
% |A(\hat{p}_h,\hat{q}_h)|\ \rangle$ as usually what we do in
% Heisenberg picture and classical quantities
% $f(\textbf{P}_h,\textbf{Q}_h)$ are also obvious as in the case of
% Schr\"{o}dinger picture.

If we hold the classical variables fixed, the hybrid system is reduced
to a fully quantum mechanical system. In this case, the mean field
theory is just the Schr\"{o}dinger picture and the quasiclassical
bracket approach becomes the Heisenberg picture. Their equivalence
has been proved a long time ago by Dirac\cite{Dirac}.  It is not clear whether
this is still true when the classical variables are allowed to evolve
under the influence of the quantum backreaction in a hybrid system.

We know that the quantum dynamics described by Eq.(\ref{sch}) is the
same as the one described by Eqs.(\ref{qqh},\ref{pph}) if we have
\begin{equation}
\label{QP}
\textbf{Q}_s(t)=\textbf{Q}_h(t)\,,~~~~
\textbf{P}_s(t)=\textbf{P}_h(t)\,.
\end{equation}
Consequently, the whole proof comes down to show that the
above equalities hold. We compare Eqs.(\ref{qs},\ref{ps})
and Eqs.(\ref{qh},\ref{ph}). The only difference is the quantum
backreaction force. One is given by Eq.(\ref{fs}) and the other by
\begin{equation}
{\bf F}_h=-\frac{\partial }{\partial
\textbf{Q}_h}\Big[\langle t_0|\hat{H}_{ih}(t)|t_0\rangle\Big]\,.
\end{equation}
Whether these two forces are the same depends on whether
\begin{equation}
 E_{s}(t)= \langle \psi|\hat{H}_{is}|\psi\rangle
 \end{equation}
 and
 \begin{equation}
 E_{h}(t)= \langle t_0|\hat{H}_{ih}(t)|t_0\rangle
 \end{equation}
 are identical.  As we shall show, we indeed have $E_s(t)=E_h(t)$.

 We notice that $\textbf{Q},\textbf{P}$ and $E$ are mutually dependent.
 That is $\textbf{Q},\textbf{P}$ depend on $E$ and at the same time
 $E$ depends on $\textbf{Q},\textbf{P}$. This mutual dependence
 means that the equalities in Eq.(\ref{QP}) are equivalent to a more
 complete set of equalities
 \begin{equation}
 \label{QPE}
\textbf{Q}_s(t)=\textbf{Q}_h(t)\,,~~
\textbf{P}_s(t)=\textbf{P}_h(t)\,,~~E_s(t)=E_h(t)\,.
 \end{equation}
 Once these equalities are proven, the proof is done. The time evolutions
 of  $E_s$ and $E_h$ are very similar.  For $E_s$,  we have
\begin{eqnarray}
\frac{dE_s(t)}{dt}
&=&\Big(\frac{d}{dt}\langle\psi|\Big)\hat{H}_{ih}|\psi\rangle
+\langle\psi|\hat{H}_{ih}\Big(\frac{d}{dt}|\psi\rangle\Big)\nonumber\\
&=&\frac{1}{i\hbar}\langle\psi|
[\hat{H}_{is},\hat{H}_{qs}+\hat{H}_{is}]|\psi\rangle\nonumber\\
&=&\frac{1}{i\hbar}\langle\psi|[\hat{H}_{is},\hat{H}_{qs}]|\psi\rangle\,.
\end{eqnarray}
For $E_h$, we get
\begin{eqnarray}
\frac{dE_h(t)}{dt}&=&\langle t_0|\frac{d}{dt}\hat{H}_{ih}(t)|t_0\rangle\nonumber\\
&=&\frac{1}{i\hbar}\langle t_0 |
[\hat{H}_{ih}(t),\hat{H}_{qh}(t)]|t_0\rangle\,.
%&=&\frac{1}{i\hbar}\langle t_0|e^{i\frac{(\hat{H}_{qh}+\hat{H}_{ih})t}{\hbar}}
%[\hat{H}_{ih},\hat{H}_{qh}]e^{-i\frac{(\hat{H}_{qh}+\hat{H}_{ih})t}{\hbar}}|t_0\rangle.
\end{eqnarray}

We are ready for the final step of our proof. At the initial moment $t_0$, we have
\begin{equation}
\dot{\bf Q}_s(t_0)=\dot{\bf Q}_h(t_0),~
\dot{\bf P}_s(t_0)=\dot{\bf P}_h(t_0),
\end{equation}
and
\begin{equation}
E_s(t_0)=E_h(t_0),~~~\frac{dE_s(t)}{dt}\Big|_{t_0}=\frac{dE_h(t)}{dt}\Big|_{t_0}\,.
\end{equation}
These equalities imply that, at the next moment $t_1=t_0+dt$, we have
\begin{align}
{\bf Q}_s(t_1)&={\bf Q}_s(t_0)+\dot{\bf Q}_s(t_0)dt\nonumber\\
                         &={\bf Q}_h(t_0)+\dot{\bf Q}_h(t_0)dt={\bf Q}_h(t_1)\,,
\end{align}
\begin{align}
{\bf P}_s(t_1)&={\bf P}_s(t_0)+\dot{\bf P}_s(t_0)dt\nonumber\\
                         &={\bf P}_h(t_0)+\dot{\bf P}_h(t_0)dt={\bf P}_h(t_1)\,,
\end{align}
and
\begin{align}
E_s(t_1)&=E_s(t_0)+\frac{dE_s(t)}{dt}\Big|_{t_0}dt\nonumber\\
                         &=E_h(t_0)+\frac{dE_h(t)}{dt}\Big|_{t_0}dt=E_h(t_1)\,.
\end{align}
For the following moments $t_2=t_1+dt$,
$t_3=t_2+dt,\cdots,t_n=t_{n-1}+dt,\cdots$, we can similarly show
that the equalities in Eq.(\ref{QPE}) hold.  This completes our
proof that the mean-field theory and the quasiclassical approach are
equivalent.

There is an  alternative to the above proof. We outline it here. One
first notice that both sets of equations, Eqs.(\ref{sch})-(\ref{ps})
and Eqs.\eqref{qqh}-\eqref{ph}, have unique solutions once the
initial conditions are specified. Then the equivalence is proved
when one shows that the solution of one set of equations also
satisfies the other set.

We have calculated numerically the example in Ref. \cite{ref:zq} with
these two different approaches and the results confirm  the above
proof.  Our experience is that the numerical computation
with the mean-field theory is much less time consuming than
the other approach.

Although quantum mechanics and classical mechanics are very different
physically, they share at least two common mathematical features. One
is that the Schr\"odinger equation has also a classical Hamiltonian
structure\cite{ref:heslot,ref:weinberg}, which is utilized by the mean-field
theory of a hybrid system. The other is that the Poisson brackets in the
classical mechanics
and the quantum commutators share a similar algebraic structure.
The common feature is explored by the quasiclassical bracket
approach.  Interestingly, these seemingly quite different methods
lead to the same dynamics as we have shown.  However, there
is a difference between these two approaches that is worth mentioning.

The mean-field theory is mathematically rigorous. One can derive the
equations of motion Eqs.(\ref{sch})-(\ref{ps}) from the Hamiltonian
in Eq.(\ref{mf})by rigorously following the classical Hamiltonian
theory. However, this is not so for the quasiclassical approach.
First, there are several different ways to setting up the
quasiclassical brackets as we have mentioned; second, Eq.(\ref{ph})
is written with some arbitrariness. There can be other alternatives.
For example, one obvious alternative is to replace $\langle
t_0|\hat{H}_{ih}|t_0\rangle$ with
\begin{equation}
H_{i}({\bf Q}_{h},\langle t_0|\hat{q}_h|t_0\rangle)\,.
\end{equation}
There seem no a priori principles that favor one over another. One can
only make a choice based on the consequence of each choice.

In conclusion, we have proved the equivalence of two popular but
different methods for quantum-classical hybrid systems. This
conclusion suggests that many approaches that have been proposed for
hybrid systems may also equivalent to one another.

\begin{acknowledgments}
We thank Yongping Zhang, Li Mao, and Wu Li for useful discussion, and
Hu Xu for supplying computational resource. B.W. acknowledges the support
of the Daniel Tsui Fellowship of the University of Hong Kong.
This work is supported by the ``BaiRen'' program of Chinese Academy
of Sciences, the NSF of China (10504040), and the 973 project of
China(2005CB724500,2006CB921400).
\end{acknowledgments}
%\bibliography{ref}

\begin{thebibliography}{36}
\expandafter\ifx\csname natexlab\endcsname\relax\def\natexlab#1{#1}\fi
\expandafter\ifx\csname bibnamefont\endcsname\relax
  \def\bibnamefont#1{#1}\fi
\expandafter\ifx\csname bibfnamefont\endcsname\relax
  \def\bibfnamefont#1{#1}\fi
\expandafter\ifx\csname citenamefont\endcsname\relax
  \def\citenamefont#1{#1}\fi
\expandafter\ifx\csname url\endcsname\relax
  \def\url#1{\texttt{#1}}\fi
\expandafter\ifx\csname urlprefix\endcsname\relax\def\urlprefix{URL }\fi
\providecommand{\bibinfo}[2]{#2}
\providecommand{\eprint}[2][]{\url{#2}}

\bibitem[{\citenamefont{Rosenfeld}(1963)}]{Rosenfeld1963NPhy}
\bibinfo{author}{\bibfnamefont{L.}~\bibnamefont{Rosenfeld}},
  \bibinfo{journal}{Nuclear Physics} \textbf{\bibinfo{volume}{40}},
  \bibinfo{pages}{353} (\bibinfo{year}{1963}).

\bibitem[{\citenamefont{Kibble}(1981)}]{Kibble1981}
\bibinfo{author}{\bibfnamefont{T.~W.~B.} \bibnamefont{Kibble}}, in
  \emph{\bibinfo{booktitle}{Quantum Gravity 2}}, edited by
  \bibinfo{editor}{\bibfnamefont{C.~J.} \bibnamefont{Isham}},
  \bibinfo{editor}{\bibfnamefont{R.}~\bibnamefont{Penrose}}, \bibnamefont{and}
  \bibinfo{editor}{\bibfnamefont{D.~W.} \bibnamefont{Sciama}}
  (\bibinfo{publisher}{Clarendon Press}, \bibinfo{address}{Oxford},
  \bibinfo{year}{1981}), p.~\bibinfo{pages}{63}.

\bibitem[{\citenamefont{{A. Anderson}}(1995)}]{ref:andsn}
\bibinfo{author}{\bibnamefont{{A. Anderson}}}, \bibinfo{journal}{Phys. Rev.
  Lett.} \textbf{\bibinfo{volume}{74}}, \bibinfo{pages}{621}
  (\bibinfo{year}{1995}).

\bibitem[{\citenamefont{Braginsky and Khalili}(1992)}]{Braginsky1992Book}
\bibinfo{author}{\bibfnamefont{V.~B.} \bibnamefont{Braginsky}}
  \bibnamefont{and} \bibinfo{author}{\bibfnamefont{F.~Y.}
  \bibnamefont{Khalili}}, \emph{\bibinfo{title}{Quantum Measurement}}
  (\bibinfo{publisher}{Cambridge}, \bibinfo{year}{1992}).

\bibitem[{\citenamefont{{L. Di\'{o}si, N. Gisin, and W. T.
  Strunz}}(2000)}]{ref:gisin}
\bibinfo{author}{\bibnamefont{{L. Di\'{o}si, N. Gisin, and W. T. Strunz}}},
  \bibinfo{journal}{Phys. Rev. A} \textbf{\bibinfo{volume}{61}},
  \bibinfo{pages}{022108} (\bibinfo{year}{2000}).

\bibitem[{\citenamefont{Tully}(1998)}]{ref:tully489}
\bibinfo{author}{\bibfnamefont{J.~C.} \bibnamefont{Tully}}, in
  \emph{\bibinfo{booktitle}{Classical and Quantum Dynamics in Condensed Phase
  Simulations}}, edited by \bibinfo{editor}{\bibfnamefont{B.~J.}
  \bibnamefont{Berne}},
  \bibinfo{editor}{\bibfnamefont{G.}~\bibnamefont{Ciccotti}}, \bibnamefont{and}
  \bibinfo{editor}{\bibfnamefont{D.~F.} \bibnamefont{Coker}}
  (\bibinfo{publisher}{World Scientific}, \bibinfo{address}{Singapore},
  \bibinfo{year}{1998}), pp. \bibinfo{pages}{489--514}.

\bibitem[{\citenamefont{{O. V. Prezhdo and V. V. Kisil}}(1997)}]{ref:pk}
\bibinfo{author}{\bibnamefont{{O. V. Prezhdo and V. V. Kisil}}},
  \bibinfo{journal}{Phys. Rev. A} \textbf{\bibinfo{volume}{56}},
  \bibinfo{pages}{162} (\bibinfo{year}{1997}).

\bibitem[{\citenamefont{{O. V. Prezhdo, and P. J.
  Rossky}}(1997)}]{ref:prezhdomean}
\bibinfo{author}{\bibnamefont{{O. V. Prezhdo, and P. J. Rossky}}},
  \bibinfo{journal}{J. Chem. Phys.} \textbf{\bibinfo{volume}{107}},
  \bibinfo{pages}{825} (\bibinfo{year}{1997}).

\bibitem[{\citenamefont{{J. B. Delos, W. R. Thorson, and S. K.
  Knudson}}(1972)}]{ref:delos1}
\bibinfo{author}{\bibnamefont{{J. B. Delos, W. R. Thorson, and S. K.
  Knudson}}}, \bibinfo{journal}{Phys. Rev. A} \textbf{\bibinfo{volume}{6}},
  \bibinfo{pages}{709} (\bibinfo{year}{1972}).

\bibitem[{\citenamefont{{J. B. Delos, and W. R. Thorson}}(1972)}]{ref:delos2}
\bibinfo{author}{\bibnamefont{{J. B. Delos, and W. R. Thorson}}},
  \bibinfo{journal}{Phys. Rev. A} \textbf{\bibinfo{volume}{6}},
  \bibinfo{pages}{720} (\bibinfo{year}{1972}).

\bibitem[{\citenamefont{Born and Oppenheimer}(1927)}]{Born1927}
\bibinfo{author}{\bibfnamefont{M.}~\bibnamefont{Born}} \bibnamefont{and}
  \bibinfo{author}{\bibfnamefont{J.~R.} \bibnamefont{Oppenheimer}},
  \bibinfo{journal}{Ann. Phys. (Leipzig)} \textbf{\bibinfo{volume}{84}},
  \bibinfo{pages}{457} (\bibinfo{year}{1927}).

\bibitem[{\citenamefont{Ehrenfest}(1927)}]{ref:eh}
\bibinfo{author}{\bibfnamefont{P.}~\bibnamefont{Ehrenfest}},
  \bibinfo{journal}{Z. Phys.} \textbf{\bibinfo{volume}{45}},
  \bibinfo{pages}{455} (\bibinfo{year}{1927}).

\bibitem[{\citenamefont{{Q. Zhang, B. Wu}}(2006)}]{ref:zq}
\bibinfo{author}{\bibnamefont{{Q. Zhang, B. Wu}}}, \bibinfo{journal}{Phys. Rev.
  Lett.} \textbf{\bibinfo{volume}{97}}, \bibinfo{pages}{190401}
  (\bibinfo{year}{2006}).

\bibitem[{\citenamefont{Aleksandrov}(1981)}]{ref:alek}
\bibinfo{author}{\bibfnamefont{I.~V.} \bibnamefont{Aleksandrov}},
  \bibinfo{journal}{Z. Nuturforsch. A.} \textbf{\bibinfo{volume}{36}},
  \bibinfo{pages}{902} (\bibinfo{year}{1981}).

\bibitem[{\citenamefont{{R. Kapral and G. Ciccotti}}(1999)}]{ref:kapral}
\bibinfo{author}{\bibnamefont{{R. Kapral and G. Ciccotti}}},
  \bibinfo{journal}{J. Chem. Phys.} \textbf{\bibinfo{volume}{110}},
  \bibinfo{pages}{8919} (\bibinfo{year}{1999}).

\bibitem[{\citenamefont{{S. Nielsen, R. Kapral, and G.
  Ciccotti}}(2001)}]{ref:nielsen}
\bibinfo{author}{\bibnamefont{{S. Nielsen, R. Kapral, and G. Ciccotti}}},
  \bibinfo{journal}{J. Chem. Phys.} \textbf{\bibinfo{volume}{115}},
  \bibinfo{pages}{5805} (\bibinfo{year}{2001}).

\bibitem[{\citenamefont{{O. V. Prezhdo and C.
  Brooksby}}(2001)}]{ref:prezhdoback}
\bibinfo{author}{\bibnamefont{{O. V. Prezhdo and C. Brooksby}}},
  \bibinfo{journal}{Phys. Rev. Lett.} \textbf{\bibinfo{volume}{86}},
  \bibinfo{pages}{3215} (\bibinfo{year}{2001}).

\bibitem[{\citenamefont{{E. Gindensperger, C. Meier, and J. A.
  Beswick}}(2000)}]{ref:ginden}
\bibinfo{author}{\bibnamefont{{E. Gindensperger, C. Meier, and J. A.
  Beswick}}}, \bibinfo{journal}{J. Chem. Phys.} \textbf{\bibinfo{volume}{113}},
  \bibinfo{pages}{9369} (\bibinfo{year}{2000}).

\bibitem[{\citenamefont{Halliwell}(1998)}]{ref:hallid}
\bibinfo{author}{\bibfnamefont{J.~J.} \bibnamefont{Halliwell}},
  \bibinfo{journal}{Phys. Rev. D} \textbf{\bibinfo{volume}{57}},
  \bibinfo{pages}{2337} (\bibinfo{year}{1998}).

\bibitem[{\citenamefont{{E. R. Bittner and P. J. Rossky}}(1995)}]{ref:bittner}
\bibinfo{author}{\bibnamefont{{E. R. Bittner and P. J. Rossky}}},
  \bibinfo{journal}{J. Chem. Phys.} \textbf{\bibinfo{volume}{103}},
  \bibinfo{pages}{8130} (\bibinfo{year}{1995}).

\bibitem[{\citenamefont{Jones}(1992)}]{ref:jonesd}
\bibinfo{author}{\bibfnamefont{K.~R.~W.} \bibnamefont{Jones}},
  \bibinfo{journal}{Phys. Rev. D} \textbf{\bibinfo{volume}{45}},
  \bibinfo{pages}{R2590} (\bibinfo{year}{1992}).

\bibitem[{\citenamefont{Tully and Preston}(1971)}]{ref:tullytrajectory}
\bibinfo{author}{\bibfnamefont{J.~C.} \bibnamefont{Tully}} \bibnamefont{and}
  \bibinfo{author}{\bibfnamefont{R.~K.} \bibnamefont{Preston}},
  \bibinfo{journal}{J. Chem. Phys.} \textbf{\bibinfo{volume}{55}},
  \bibinfo{pages}{562} (\bibinfo{year}{1971}).

\bibitem[{\citenamefont{Tully}(1990)}]{ref:tullymolecular}
\bibinfo{author}{\bibfnamefont{J.~C.} \bibnamefont{Tully}},
  \bibinfo{journal}{J. Chem. Phys.} \textbf{\bibinfo{volume}{93}},
  \bibinfo{pages}{1061} (\bibinfo{year}{1990}).

\bibitem[{\citenamefont{{A. Heslot}}(1985)}]{ref:heslot}
\bibinfo{author}{\bibnamefont{{A. Heslot}}}, \bibinfo{journal}{Phys. Rev. D}
  \textbf{\bibinfo{volume}{31}}, \bibinfo{pages}{1341} (\bibinfo{year}{1985}).

\bibitem[{\citenamefont{Weinberg}(1989)}]{ref:weinberg}
\bibinfo{author}{\bibfnamefont{S.}~\bibnamefont{Weinberg}},
  \bibinfo{journal}{Ann. Phys. (N. Y.)} \textbf{\bibinfo{volume}{194}},
  \bibinfo{pages}{336} (\bibinfo{year}{1989}).

\bibitem[{\citenamefont{{J. Liu, B. Wu, and Q. Niu}}(2003)}]{ref:lj}
\bibinfo{author}{\bibnamefont{{J. Liu, B. Wu, and Q. Niu}}},
  \bibinfo{journal}{Phys. Rev. Lett.} \textbf{\bibinfo{volume}{90}},
  \bibinfo{pages}{170404} (\bibinfo{year}{2003}).

\bibitem[{\citenamefont{Epstein}(1981)}]{ref:ep}
\bibinfo{author}{\bibfnamefont{S.~T.} \bibnamefont{Epstein}}, in
  \emph{\bibinfo{booktitle}{Force Concept in Chemistry}}
  (\bibinfo{publisher}{Van Nostrand Reinhold}, \bibinfo{address}{New York},
  \bibinfo{year}{1981}), pp. \bibinfo{pages}{1--38}.

\bibitem[{\citenamefont{{S. Hammes-Schiffer and J. C.
  Tully}}(1994)}]{ref:hammes}
\bibinfo{author}{\bibnamefont{{S. Hammes-Schiffer and J. C. Tully}}},
  \bibinfo{journal}{J. Chem. Phys.} \textbf{\bibinfo{volume}{101}},
  \bibinfo{pages}{4657} (\bibinfo{year}{1994}).

\bibitem[{\citenamefont{{L. Di\'{o}si and J. J. Halliwell}}(1998)}]{ref:dh}
\bibinfo{author}{\bibnamefont{{L. Di\'{o}si and J. J. Halliwell}}},
  \bibinfo{journal}{Phys. Rev. Lett.} \textbf{\bibinfo{volume}{81}},
  \bibinfo{pages}{2846} (\bibinfo{year}{1998}).

\bibitem[{\citenamefont{{V. V. Kisil}}(2005)}]{ref:kisil}
\bibinfo{author}{\bibnamefont{{V. V. Kisil}}}, \bibinfo{journal}{Europhys.
  Lett} \textbf{\bibinfo{volume}{72}}, \bibinfo{pages}{873}
  (\bibinfo{year}{2005}).

\bibitem[{\citenamefont{{O. V. Prezhdo}}(2006)}]{ref:prezhdoquantum}
\bibinfo{author}{\bibnamefont{{O. V. Prezhdo}}}, \bibinfo{journal}{J. Chem.
  Phys.} \textbf{\bibinfo{volume}{124}}, \bibinfo{pages}{201104}
  (\bibinfo{year}{2006}).

\bibitem[{\citenamefont{{W. Boucher and J. Traschen}}(1988)}]{ref:boucher}
\bibinfo{author}{\bibnamefont{{W. Boucher and J. Traschen}}},
  \bibinfo{journal}{Phys. Rev. D} \textbf{\bibinfo{volume}{37}},
  \bibinfo{pages}{3522} (\bibinfo{year}{1988}).

\bibitem[{\citenamefont{{L. Di{\'o}si}}(1996)}]{ref:diosi}
\bibinfo{author}{\bibnamefont{{L. Di{\'o}si}}}, \bibinfo{journal}{Phys. Rev.
  Lett.} \textbf{\bibinfo{volume}{76}}, \bibinfo{pages}{4088}
  (\bibinfo{year}{1996}).

\bibitem[{\citenamefont{{L. L. Salcedo}}(2007)}]{ref:salcedo}
\bibinfo{author}{\bibnamefont{{L. L. Salcedo}}}, \bibinfo{journal}{J. Chem.
  Phys.} \textbf{\bibinfo{volume}{126}}, \bibinfo{pages}{057101}
  (\bibinfo{year}{2007}).

\bibitem[{\citenamefont{{K. R. W. Jones}}(1996)}]{ref:jones}
\bibinfo{author}{\bibnamefont{{K. R. W. Jones}}}, \bibinfo{journal}{Phys. Rev.
  Lett.} \textbf{\bibinfo{volume}{76}}, \bibinfo{pages}{4087}
  (\bibinfo{year}{1996}).

\bibitem[{\citenamefont{Dirac}(1967)}]{Dirac}
\bibinfo{author}{\bibfnamefont{P.~A.~M.} \bibnamefont{Dirac}},
  \emph{\bibinfo{title}{the Principles of Quantum Mechanics}}
  (\bibinfo{publisher}{Clarendon Press}, \bibinfo{address}{Oxford},
  \bibinfo{year}{1967}), \bibinfo{edition}{4th} ed.

\end{thebibliography}

\end{document}